\documentclass[twocolumn,showpacs,preprintnumbers,amsmath,amssymb,prl,epsf]{revtex4}
\usepackage{graphicx}
\begin{document}
\title{Entanglement generation via phase-matched processes: different Bell states within the linewidth}
\author{G. Brida, M. Genovese, L.~A.~Krivitsky}
\affiliation{Istituto Nazionale di Ricerca Metrologica,\\
Strada delle Cacce 91, 10135 Torino, Italy}
\author{M.~V.~Chekhova }
\affiliation{Department of Physics, M.V.Lomonosov Moscow State
University,\\  Leninskie Gory, 119992 Moscow, Russia} \vskip 24pt

\begin{abstract}
\begin{center}\parbox{14.5cm}
{It is shown, theoretically and experimentally, that at any type-II
spontaneous parametric down-conversion (SPDC) phase matching, the
decoherence-free singlet Bell state is always present within the
natural bandwidth and can be filtered out by a proper spectral
selection. Instead of the frequency selection, one can perform time
selection of the two-photon time amplitude at the output of a
dispersive fibre. Applications to quantum communication are
outlined.}
\end{center}
\end{abstract}
 \maketitle \narrowtext
\vspace{-10mm}


Type-II spontaneous parametric down-conversion (SPDC) provides the
easiest way of generating polarization-entangled two-photon states,
which find applications  in quantum communication and quantum
computation. When using polarization-entangled two-photon states in
quantum communication protocols, the basic problem is their
decoherence \cite{gis} due to the propagation of two-photon light
through optical fibres or even through free space \cite{ant}, if one
takes into account the effect of the atmosphere. However, it has
been shown~\cite{Kwiatscience,Dragan} that decoherence-free
\cite{ras} propagation of polarization-entangled two-photon light
can be achieved by using the singlet Bell state, $\Psi^-$, which is
invariant to any polarization transformation.

Robustness against decoherence is not the only remarkable property
of the singlet two-photon Bell state. Being antisymmetric with
respect to the permutation of the two photons, it can be filtered
out of the set of Bell states with the help of a 50\% non-polarizing
beamsplitter, due to the Hong-Ou-Mandel 'dip' effect~\cite{filter}.
Because of this property, it was the basic tool in the first quantum
teleportation experiments~\cite{teleportation}. This state is
unpolarized not only in the usual sense (considering the intensity
moments) but also with respect to all higher-order intensity
moments~\cite{Yura}. Its 'bright' (high-photon-number) analogue is
predicted to have the properties of 'scalar light', for which the
fluctuations of all Stokes parameters are suppressed below the
shot-noise limit~\cite{KarasMas}.

There are two basic types of polarization-entangled singlet
two-photon state \cite{prep}. Historically, the first method to
efficiently prepare the singlet state was based on non-collinear
type-II SPDC~\cite{Kwiat_rings}. Later, it was suggested to produce
the same state by using two type-I SPDC crystals under non-collinear
phase matching~\cite{Kwiat}. In both these cases, the state $\Psi^-$
is prepared in two different spatial modes and can be called a
polarization-wavevector entangled state. The second type of a
singlet state is a polarization-frequency entangled state, that is,
a polarization-entangled state in two different frequency modes.
This can be done by using two type-I SPDC crystal under collinear
frequency-nondegenerate phasematching~\cite{Sergei+Yoon-Ho,Yura}.
For quantum communication applications this type of the singlet
state is preferable since both photons of an entangled pair can be
easily transmitted through the same optical fibre and undergo the
same polarization changes.

It is important to point out that to achieve polarization
entanglement, the schemes based on type-II SPDC are always
provided with a birefringent crystal compensating the e-o delay
$\tau_0$ between the orthogonally polarized photons of a single
pair. Alternatively, one can use interferometric schemes based on
two type-I or two type-II crystals~\cite{two_UMBC}.

In this paper we show that the o-e delay $\tau_0$ between the signal and idler photons, which
 is caused by the nonlinear crystal and requires certain efforts to be
eliminated, can be actually quite helpful. Namely, it turns out that
the state $\Psi^-$ is always produced within the natural bandwidth
of type-II frequency-degenerate SPDC, and it is the $\tau_0$ delay
that is responsible for creating $\Psi^-$. This fact, beyond
increasing the understanding of generation and manipulation of
optical entangled states, can have important applications to quantum
communication.


Consider collinear frequency-degenerate type-II SPDC from a
continuous wave (cw) pump. In the low-gain regime the state of the
two-photon light generated this way can be represented as a
superposition of a vacuum state and a two-photon state given by the
integral over the SPDC frequency spectrum,

\begin{eqnarray}
|\Psi\rangle=|\hbox{vac}\rangle+\int{d\Omega}F(\Omega)[a^{\dagger}_{H}(\omega_0+\Omega)
a^{\dagger}_{V}(\omega_0-\Omega)e^{i\Omega\tau_0}
\nonumber\\
+a^{\dagger}_{V}(\omega_0+\Omega)
a^{\dagger}_{H}(\omega_0-\Omega)e^{-i\Omega\tau_0}]|vac\rangle,
\label{mainstate}
\end{eqnarray}
where $\omega_0=\omega_p/2$, $\omega_p$ is the pump frequency,
$a^{\dagger}_{H}$ and  $a^{\dagger}_{V}$ are the photon creation
operators in the horizontal and vertical polarization modes
(denoted by $H,V$). Since orthogonally polarized photons have
different group velocities when propagating through the crystal,
the phase factor $e^{\pm{i\Omega\tau_0}}$ appears, where
$\tau_0={DL/2}$ is the mean temporal delay between orthogonally
polarized photons, $D\equiv{1/u_V-1/u_H}$ is the difference of the
inverse group velocities and $L$, the length of the crystal. The
spectral amplitude of the state has the form

\begin{equation}
F(\Omega)=\frac{\sin(\Omega\tau_0)}{\Omega\tau_0}. \label{sinc}
\end{equation}

In existing experiments with frequency-degenerate SPDC and frequency
selection, it is always the frequency corresponding to exact
degeneracy, $\Omega=0$, that is selected. The resulting two-photon
state,  a factorized one,
$a^{\dagger}_{H}(\omega_0)a^{\dagger}_{V}(\omega_0)$,  can be turned
into $\Psi^+$ by splitting the beam on a beamsplitter~\cite{lost}.
However, if one considers a small frequency shift from the exact
degeneracy condition, $\Omega=\pi/2\tau_0$, then the two-photon part
of the state (\ref{mainstate}) is
$$
\Psi^-\equiv F(\pi/2\tau_0)[a^{\dagger}_{H}(\omega_1)
a^{\dagger}_{V}(\omega_2)-a^{\dagger}_{V}(\omega_1)
a^{\dagger}_{H}(\omega_2)],
$$
where $\omega_1=\omega_0-\pi/2\tau_0$
and $\omega_2=\omega_0+\pi/2\tau_0$. The square modulus of the
two-photon amplitude, which gives the total number of photon pairs,
is in this case $0.41$; it means that the singlet Bell state is
produced with the efficiency almost twice higher than the $\Psi^+$
state is produced using a filter and a beamsplitter~\cite{lost}.
Therefore, the singlet Bell state is present within the natural
bandwidth of any type-II SPDC spectrum and can be filtered out using
a proper spectral selection.


In our experiment we demonstrate the generation of the $\Psi^-$
state on the slopes of type-II SPDC spectral line by selecting the
frequency of one of the photons using a monochromator. The setup is
shown in Fig.1. Two-photon light was generated via spontaneous
parametric down-conversion by pumping a type-II 0.5 mm
$\beta$-barium borate crystal (BBO) with 0.45 Watt $\hbox{Ar}^{+}$
cw laser beam at the wavelength 351 nm in the collinear
frequency-degenerate regime. It is important that no birefringent
material was inserted after the crystal to compensate for the e-o
delay. The pump laser beam was eliminated by a highly reflecting UV
mirror and the SPDC radiation was addressed to a 50/50
non-polarizing beamsplitter. To perform polarization selection, two
Glan prisms were placed at the output ports of the beamsplitter. The
spectral distribution of the coincidences was analyzed with a
diffractive-grating monochromator with the resolution 0.8 nm placed
in one of the output ports of the beamsplitter. Since the pump was
cw, frequency selection in one output port automatically selected
the frequency of the correlated photon. In order to reduce the
contribution of accidental coincidences, a broadband interference
filter centered around 702 nm was placed after the beamsplitter. Its
transmission band (FWHM=40nm) was wider than the natural width of
the SPDC spectrum under given experimental conditions (FWHM=$12$nm).
Biphoton pairs were registered by two photodetection apparatuses,
consisting of red-glass filters, pinholes, focusing lenses and
avalanche photodiodes (Perkin\&Elmer single-photon counting
modules). The photocount pulses of the two detectors, after passing
through delay lines, were sent to the START and STOP inputs of a
Time--to--Amplitude Converter (TAC). The output of the TAC was
finally addressed to a Multi--Channel Analyzer (MCA), and the number
of coincidences of photocounts of the two detectors was observed at
the MCA output.


First, we studied the dependence of the coincidences counting rate
on the wavelength selected by the monochromator. In  the absence of
the Glan prisms, we obtained the usual type-II SPDC spectrum with a
FWHM of $12$ nm. If two Glan prisms oriented at angles $\theta_1,
\theta_2$ are inserted in the beamsplitter output ports, the
coincidence counting rate $R_c$ should depend on the selected
frequency offset $\Omega$ from exact degeneracy as~\cite{PRAfibre}

\begin{eqnarray}
R_c=\frac{\sin^2(\Omega\tau_0)}{(\Omega\tau_0)^2}[\sin^2(\theta_1+\theta_2)\cos^2(\Omega\tau_0)
\\
\nonumber +\sin^2(\theta_1-\theta_2)\sin^2(\Omega\tau_0)].
\label{thetas}
\end{eqnarray}

Typical dependencies of $R_c$ on $\Omega$ at $\theta_1=\pi/4$ and
different values of $\theta_2$ are shown in Fig.2. This behaviour
demonstrates two-photon interference, which in this case manifests
itself within the lineshape of SPDC frequency spectrum, as it was
indeed predicted in Ref.~\cite{PRAfibre}.

The interference is observed most clearly
 for two orientations of the Glan prism in channel 2: at
$\theta_2=45^\circ$ and at $\theta_2=-45^\circ$. The experimental
dependencies obtained for these cases are shown in Fig.3: in perfect
agreement with formula (\ref{thetas}), a maximum is observed at the
center of the spectrum for the ($45^{\circ},45^{\circ}$)
orientations of the Glan prisms and a minimum, for ($45^{\circ},
-45^{\circ}$) orientations.

From the theoretical calculation (Fig.2), as well as from the
experimental spectra (Fig.3), one can see that with certain
wavelengths selected, polarization interference with high visibility
is observed under the rotation of the Glan prism in channel 2. This
fact is well-known for the case where the selected wavelength is the
central one. However, from Fig.3 we deduce that high-visibility
polarization interference also takes place when the selected
wavelength is $695.5$ nm or $708.5$ nm. Both cases correspond to the
selection of the $|\Psi^-\rangle$ state.

To confirm this fact experimentally, we have measured the
polarization interference for the $|\Psi^-\rangle$ state: the
wavelength transmitted by the monochromator was fixed at $708.5$ nm,
and the coincidence counting rate was measured depending on the
orientation of one of the polarizers, the other polarizer being
oriented at $\theta_1=45^\circ$. The dependence, shown in Fig.4,
demonstrates a visibility of 98\%.

In order to prove the invariance of the produced $|\Psi^-\rangle$
state under polarization transformation of the measurement basis we
have placed quarter- and half- wave plates (QWP and HWP) at various
orientations in front of the beamsplitter. In particular, if a HWP
is placed in front of the beamsplitter and its orientation is
changed, the dependencies similar to those shown in Fig.2 transform
as shown in Fig.5(a,b), where both the theoretical curve and the
experimental data are presented. As expected from the theoretical
prediction, the experimental  data prove that the coincidence
counting rate corresponding to the wavelengths $695.5$ nm and
$708.5$ nm (the $|\Psi^-\rangle$ state) both at the positions of the
Glan prisms ($45^{\circ},45^{\circ}$) and ($45^{\circ},-45^{\circ}$)
does not change depending on the HWP orientation, i.e., depending on
the rotation of the biphoton polarization state before the
beamsplitter. At all other wavelengths (including the central one,
$702$ nm), the coincidence counting rate clearly depends on the HWP
orientation. Similar behaviour is observed for a QWP inserted before
the beamsplitter.

The conclusion is that the frequency spectrum of collinear
frequency-degenerate type-II SPDC contains, at about half-width from
the center, the decoherence-free singlet Bell state
$|\Psi^-\rangle$. In order to use this state in quantum
communications, one should select a relatively narrow frequency
bandwidth. In the above-described experiment, a $0.8$ nm bandwidth
was sufficient for the case of a $0.5$ mm crystal, but using a
thicker crystal would require a more narrow frequency selection.
This means a certain experimental difficulty, which, however, can be
easily overcome if the produced state is to be transmitted through
optical fibres.

Indeed, as it was shown in Refs.~\cite{Brida,PRAfibre}, when a
two-photon state is transmitted through an optical fibre, due to the
group-velocity dispersion (GVD) the shape of the spectral amplitude
is transferred into the shape of the two-photon time amplitude and
hence, into the distribution of the time interval between the
arrivals of two photons of a pair. The frequency argument of the
spectral two-photon amplitude will then be transformed into the time
argument of the time two-photon amplitude as
$\Omega\rightarrow\tau=2k''z\Omega$, where $k''$ is the second
derivative of the dispersion law and $z$ is the fibre length.
Therefore, the frequency selection of the $|\Psi^-\rangle$ state can
be performed through the time selection of the delay between
registering two photons. This can be done by selecting events
corresponding to certain MCA channels. For instance, if a $1$ km of
single-mode fibre with $k''=3.2\cdot 10^{-28}\hbox{s}^2/\hbox{cm}$
is inserted before the beamsplitter, selection of the
$|\Psi^-\rangle$ state is provided by choosing coincidence events
with the delay between the signal and idler photons being $\pm 3$
ns~\cite{PRAfibre}.

This fact can be used in quantum communications whenever propagation
of polarization-entangled photons through optical fibres is
involved. On the one hand, the receiver can always post-select the
singlet state by picking only those coincidence events for which the
signal and idler photons come with a fixed nonzero delay $\tau=\pi
k'' z/\tau_0$. On the other hand, polarization drift introduced by
the fibre will never influence the entangled state $|\Psi^-\rangle$.

Finally, we would like to notice that in order to implement any
protocol of information transmission \cite{gis}, one should be able
to generate another entangled state at the same frequencies as the
singlet state $|\Psi^-\rangle$. In particular, the state
$|\Psi^+\rangle$ can be easily created at the same frequencies by
introducing after the crystal a birefringent material {\it twice
increasing} the $\tau_0$ delay. This can be done by inserting a
quartz plate with the same thickness as it is necessary to
compensate for the $\tau_0$ delay, but with the optic axis oriented
orthogonally \footnote{On the other hand, $\phi^{\pm}$ could be
generated by applying a polarization rotation on one arm after beam
splitting.}.

This work has been supported by MIUR (FIRB RBAU01L5AZ-002, PRIN
2005023443-002 ), by Regione Piemonte (E14), and by "San Paolo
foundation", M.Ch. also acknowledges the support of the Russian
Foundation for Basic Research, grant no.06-02-16393.

\begin{figure}
\includegraphics[height=3cm]{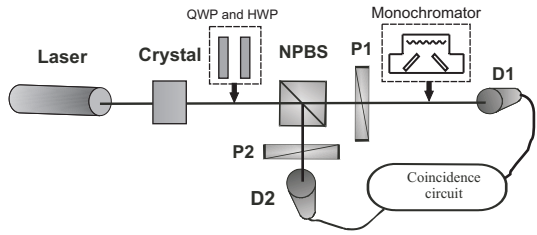} \caption{Experimental setup for
observing two-photon polarization interference within the lineshape
of the two-photon spectral amplitude. A type-II BBO crystal cut for
collinear frequency-degenerate phasematching is pumped by cw $Ar^+$
laser at 351 nm; NPBS is a 50/50 nonpolarizing beamsplitter; P1 and
P2 are linear polarization filters (Glan prisms); D1, D2 are
single-photon counting modules with outputs connected to the
coincidence counter. A diffraction-grating monochromator is placed
in one arm for the frequency selection; retardation plates (QWP and
HWP) are used to study the invariance of the Bell state
$|\Psi^-\rangle$ under polarization transformations.}
\end{figure}

\begin{figure}
\includegraphics[height=6cm]{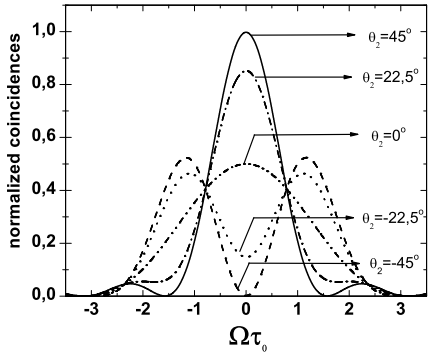} \caption{Theoretical dependence of the coincidences counting
rate on the parameter $\Omega\tau_0$ for various orientations
$\theta_2$ of the Glan prism in channel 2: ($0^{\circ},\pm
22.5^{\circ},\pm 45^{\circ}$). The Glan prism in channel 1 is fixed
at $\theta_1=45^{\circ}$.}
\end{figure}

\begin{figure}
\includegraphics[height=5cm]{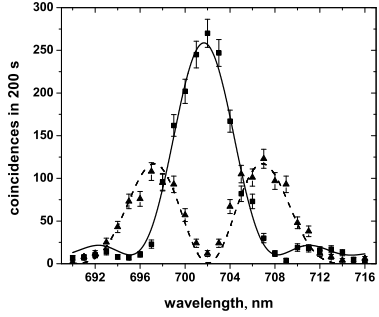}
\caption{Experimental dependence of the coincidence counting rate on
the wavelength selected by the monochromator for two cases:
$\theta_1=\theta_2=45^\circ$ (squares, solid line) and
$\theta_1=45^\circ, \theta_2=-45^\circ$ (triangles, dashed line).
The lines represent the theoretical fit to the experimental data.}
\end{figure}

\begin{figure}
\includegraphics[height=5cm]{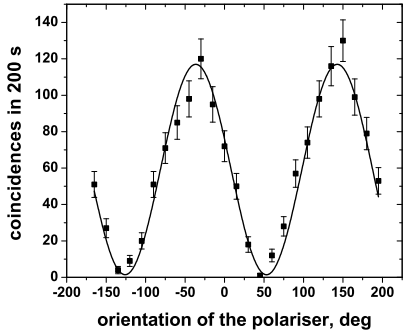}
\caption{Polarization interference fringes for the singlet Bell
state $|\Psi^-\rangle$ (the selected wavelength is
$\lambda=708.5$nm). Solid line represents the theoretical fit to the
experimental data.}
\end{figure}

\begin{figure}
\begin{tabular}{cc}
\includegraphics[width=0.3\textwidth]{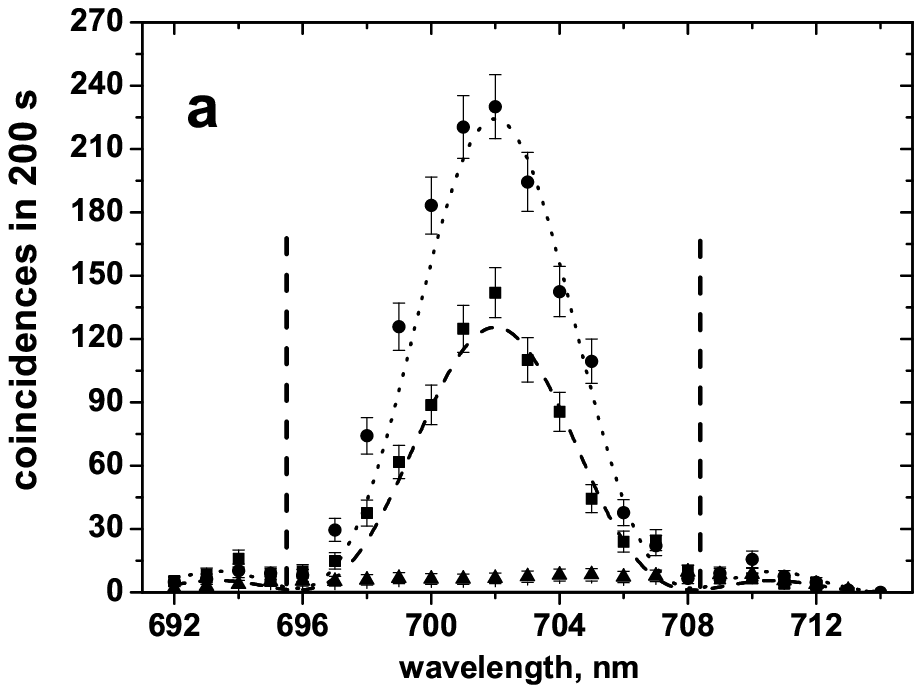}
\includegraphics[width=0.3\textwidth]{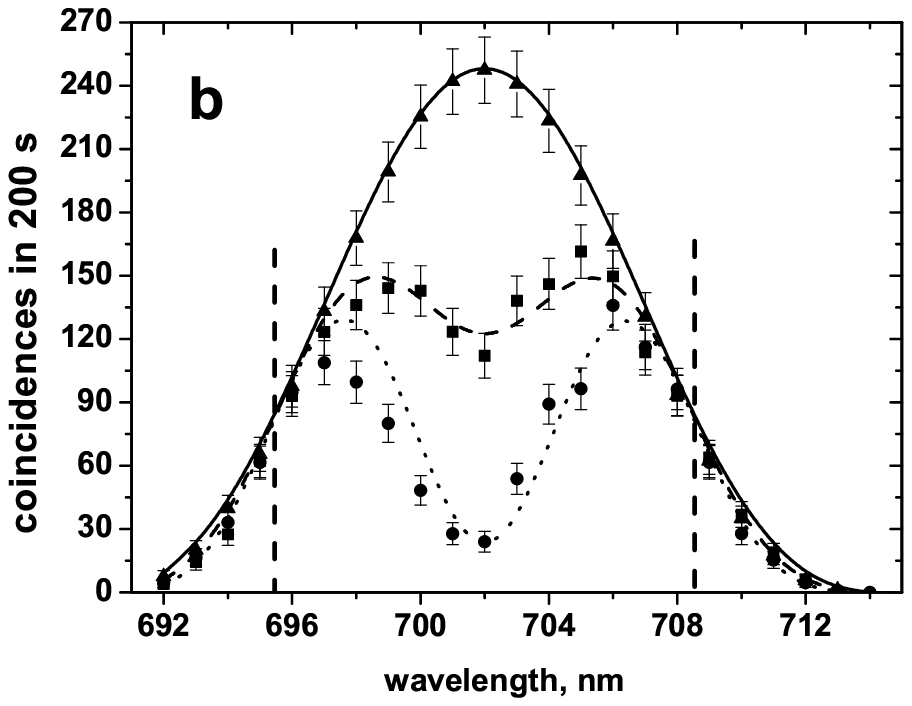}
 \end{tabular}
\caption{Experimental dependence of the coincidence counting rate on
the wavelength selected by the monochromator for the following
orientations of the HWP placed after the crystal: $7^\circ$(circles,
dotted line); $17^\circ$ (squares, dashed line ); $22,5^\circ$
(triangles, solid line). The orientations of the Glan prisms being
$45^\circ,45^\circ$ (a) and $45^\circ,-45^\circ$ (b). The high
visibility of polarization interference corresponding to the
wavelengths $\lambda=695.5$nm and $\lambda=708.5$nm (highlighted
with vertical dashed bars) confirms the invariance of the
$|\Psi^-\rangle$ state to polarization rotation. The lines represent
the theoretical fit to the experimental data.}
\end{figure}


\begin{references}

\bibitem{gis} N. Gisin et al., Rev. Mod. Phys. 74 (02) 145.
\bibitem{ant}  N. Antonietti, M. Mondin, G. Brida, M. Genovese,
quant-ph/0609049 and Ref.s therein.

\bibitem{Kwiatscience} P.G.Kwiat, A.J.Berglund, J.B.Altepeter, A.G.White. Science \textbf{290},
498-501 (2000).

\bibitem{Dragan} K.Banaszek, A.Dragan, W.Wasilewski, and Czeslaw Radzewicz. Phys. Rev. Lett. \textbf{92},
257901 (2004).
\bibitem{ras} P.Zanardi  and M.Rasetti, Phys. Rev. Lett. \textbf{79} (1997) 3306;
\bibitem{filter} S.Braunstein and A.Mann. Phys. Rev. A, \textbf{51}, R1727 (1997);
M.W.Mitchell, C.W.Ellenor, S.Schneider, and A.M.Steinberg, Phys. Rev. Lett. \textbf{91},
120402 (2003).

\bibitem{teleportation} D.Bouwmeester, J.-W.Pan, K.Mattle, M.Eibl, H.Weinfurter, and A.Zeilinger.
Nature \textbf{290}, 575-578 (1997); D.Boschi, S.Branca, F.De
Martini, L.Hardy, and S. Popescu. Phys. Rev. Lett., \textbf{80},
1121-1125 (1998).

\bibitem{Yura} A.V.Burlakov, S.P.Kulik, G.O.Rytikov, and M.V.Chekhova, JETP \textbf{95},
639-644 (2002).

\bibitem{KarasMas} V.P.Karassiov, J.Soviet Laser Res. \textbf{12},
147 (1991); V.P.Karassiov, A.V.Masalov, Opt. Spectrosc. \textbf{74},
551 (1993).
\bibitem{prep} M. Genovese,  \emph{Phys. Rep}. \textbf{413}
319 (2005).
\bibitem{Kwiat_rings} A. Garuccio, in "Fundamental Problems in Quantum Theory", Ed. D. Greenberger ,
(New York Academy of Sciences, 1995); P.G. Kwiat et al., Phys. Rev.
Lett. \textbf{75}, 4337 (1995).

\bibitem{Kwiat} L. Hardy, Phys. Lett. A 161 (1992) 326. P.G.Kwiat, E.Waks, A.G.White, I.Appelbaum, and P.G.Eberhard,
Phys. Rev. A \textbf{60}, R773 (1999). G. Brida, M. Genovese, C.
Novero and E. Predazzi, Phys. Lett. A 268, 12 (2000).

\bibitem{Sergei+Yoon-Ho} Y.H.Kim, S.P.Kulik, and Y.H.Shih. Phys. Rev. A \textbf{63},
060301(R) (2001).

\bibitem{two_UMBC} Y.-H.Kim, S.P.Kulik, M.V.Chekhova, M.H.Rubin, and Y.H.Shih. Phys. Rev. A
\textbf{63} 062301 (2001).

\bibitem{lost} Because in half of the cases, both photons go to the same output, the
state is produced with 50\% loss.

\bibitem{Brida} G. Brida, M.V. Chekhova, M. Genovese, M. Gramegna, and L.A.
Krivitsky, Phys. Rev. Lett. \textbf{96}, 143601 (2006).

\bibitem{PRAfibre} G. Brida,  M. Genovese,  L.A.
Krivitsky, and M.V. Chekhova, quant-ph 0607137, submitted to Phys.
Rev. A.


\end{references}
\end{document}